\documentclass[conference]{IEEEtran}
\IEEEoverridecommandlockouts

\usepackage{xcolor}
\usepackage{subfigure} 
\usepackage{comment}

\usepackage{hyperref}
\usepackage{url}
\usepackage{graphics}
\usepackage{graphicx}
\usepackage{balance}
\usepackage{epsfig}

\def\BibTeX{{\rm B\kern-.05em{\sc i\kern-.025em b}\kern-.08em
    T\kern-.1667em\lower.7ex\hbox{E}\kern-.125emX}}

\newcommand*{\email}[1]{#1}
\newcommand*{\affaddr}[1]{\textit{#1}}

\newcommand{\ignore}[1]{}

\begin{document}

\title{
The Metaverse Data Deluge: What Can We Do About It?
}


\author{
Beng Chin Ooi$^1$   \hspace{0.5mm} 
Gang Chen$^2$ \hspace{0.5mm} 
Mike Zheng Shou$^1$ \hspace{0.5mm} 
Kian-Lee Tan$^1$ \hspace{0.5mm} \\
Anthony Tung$^1$  \hspace{0.5mm}  
Xiaokui Xiao$^1$ \hspace{0.5mm}
James Wei Luen Yip$^3$ \hspace{0.5mm}
Meihui Zhang$^4$\\ \\
       \affaddr{$^1$National University of Singapore} \hspace{0.5mm}
       \affaddr{$^2$Zhejiang University}  \\
       \affaddr{$^3$National University Hospital}  \hspace{0.5mm} 
       \affaddr{$^4$Beijing Institute of Technology} \\
       \\
       \email{\{ooibc, tankl, atung, xiaoxk\}@comp.nus.edu.sg} \hspace{1mm} \email{cg@zju.edu.cn} \\ \email{mikeshou@nus.edu.sg} \hspace{1mm}
       \email{james\_yip@nuhs.edu.sg}
       \email{meihui\_zhang@bit.edu.cn}\\
}

\maketitle
\begin{abstract}
In the metaverse
the physical space and the virtual
space co-exist, and interact simultaneously. While the physical
space is virtually enhanced with information, the virtual space is
continuously refreshed with real-time, real-world information. To
allow users to process and manipulate information seamlessly
between the real and digital spaces, novel technologies must be
developed. These include smart interfaces, new augmented
realities, and efficient data storage, management, and dissemination
techniques. In this paper, we first discuss some promising
co-space applications. 
These applications offer
opportunities that neither of the spaces can realize on its own.
We then discuss challenges.
Finally, we discuss and envision what are likely to be required from the database and system perspectives.

\end{abstract}





\section{Introduction}

Traditionally, the physical space and the virtual space are disjoint and
distinct. Users in each space operate within the scope of the space,
i.e., they may communicate among themselves but do not cross the
boundary to the other space.
However, technological advances in hardware (e.g., GPUs, FPGAs, NVM), 
Internet of Things (IoT), 5G \cite{ooi5g}, AI, ubiquitous computing, smart interfaces, 
and new augmented realities have made it possible for these two spaces
to co-exist within a unified space, namely, the {\it metaverse}.
In a recent investment report~\cite{citi2022}, it is projected that
"the metaverse represents a potential \$8 trillion to \$13 trillion opportunity by 2030, that could boast as many as 5 billion users."
Companies such as Meta 
and Microsoft are already investing a significant amount of resources on the
metaverse. 

\begin{figure}
\centering
\includegraphics[width=2.8in]{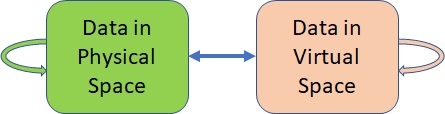}
\caption{Data Flow within the Metaverse as a result of Simultaneous Interaction}
\label{fig:co-space}
\end{figure}

In the metaverse
(or {\it cyber-physical system}), the physical space and the virtual space
interact simultaneously in real-time.
Locations and events in the physical
world are captured through the use of large numbers of sensors
and mobile devices, and could be
materialized within a virtual world. Correspondingly, certain
actions or events within the virtual domain can affect the
physical world (e.g., shopping or product promotion and experiential
computer gaming). Thus, on one hand, the physical space
is virtually enhanced with information; on the other hand, the virtual space
is continuously refreshed with real-time, real-world inputs.
Figure~\ref{fig:co-space} shows the information flow within a co-space
environment: data flow within a single space, but more importantly,
data also flow into the other space.
This distinguishes the metaverse from the mixed reality (or augmented
reality or augmented virtuality) \cite{mr20}: the mixed
reality integrates the real and virtual worlds
(e.g., augmenting live video imagery with
computer generated graphics), but it is done in a rigid and static manner,
without capturing real-time changes and their effects on either of the
spaces.
By design, the metaverse will also be more interactive and immersive than digital twins~\cite{digitaltwin19} and co-space~\cite{cospace09}. 

In the metaverse, we can design innovative applications that provide
experiences and opportunities that neither the physical nor the
virtual spaces can offer. Some example applications include location-based games and social networking, partnership in shopping among online and physical shoppers, and enhanced digital models that capture physical troop movements. 

Within such a context, it is easy to see that a large amount of
data and information must flow within the metaverse and
between the real and virtual spaces to ensure
that the two worlds are synchronized. This brings new
challenges such as a need to process heterogeneous data streams to materialize real-world events in the virtual world, and to disseminate information about interesting events in the virtual space to users in the physical world. In addition, new technologies are needed to enable users to manage information seamlessly 
between the real and digital spaces. These include smart interfaces, new augmented
realities, efficient storage management, and large-scale data dissemination. 

In this paper, we present a number of promising data-rich metaverse applications for which we believe that the database 
community has much to offer 
to drive the growth of the metaverse field.
Then, we identify several challenges that we, as data system researchers and engineers, can contribute to managing the large amount of data, the huge number
of events, and the massive number of concurrent users within
the metaverse. 
On one hand, there is a need to re-examine conventional research themes such as data fusion, data streaming, data storage and indexing,
data processing engines, distributed architectures, and data privacy and security, for the metaverse; 
on the other hand, we have to embrace new research directions to develop novel technologies that facilitate intelligent decision making, and cope with AR/VR data streaming and learning. 

The organization of the paper is as follows.
In Section~\ref{sec:datarich}.
we present five data-rich metaverse applications as examples to drive the discussion on the required database supports. 
In Section~\ref{sec:challenges}, we describe some key characteristics engendered by metaverse applications.
We discuss challenges and opportunities given rise by the metaverse in Section~\ref{sec:detailchallenges}.
We conclude in Section~\ref{sec:conclusions}.



\section{Data-rich Metaverse Use Cases}
\label{sec:datarich}

The co-existence of the physical and digital spaces offers
opportunities for novel applications. In what follows, we shall highlight
five
of these that are data-rich.

\begin{figure}[h]
\centering
\includegraphics[width=2.8in]{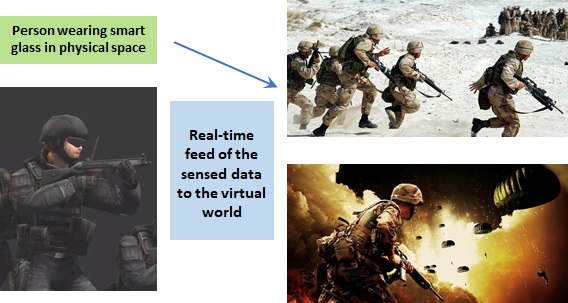}
\caption{Military Exercise and Modern War}
\label{fig:military}
\end{figure}

\subsection*{Military Mission Exercises}
Traditionally, military exercises are carried out in either the
physical realm or the virtual domain. In the physical realm, soldiers
and military vehicles are mobilized for operations in some physical
terrain. In the virtual domain, commanders ``sweat out'' in air-conditioned
rooms in a simulated warfare over 3D virtual world models of the
physical entities.
While the former is realistic, it is limited by scale (in terms of both the number of
personnel and the size of the physical space); the latter, however,
handles large-scale warfare at the expense of actual ground happenings. For example, it may take much longer time to cross a river physically than
estimated in a model because of ground constraints and the fitness of the
soldiers. 
Moreover, in a model, soldiers can walk through a building
destroyed by artillery with ease, whereas in the physical-space exercise, it may take much longer time to bypass the building.

With the metaverse, we can now conduct a more realistic
military exercise that takes on a completely new experience and
flavor. Consider an exercise that involves both a small scale
military exercise (in the physical space) and a virtual model of a
large scale military exercise. The physical space essentially
forms a small part of the entire military exercise (e.g., a
physical exercise over a physical space of 5~km by 5~km compared
to a virtual model that simulates a war over 100~km by 100~km
space). Now, on the physical ground itself, soldiers and vehicles
are equipped with location tracking devices to monitor their
movement as well as other information such as fire-power,
casualties, etc. Then, at the command center, based on real-time feed of
the sensed data, the virtual model can accurately reflect the
ground situations. Moreover, actions taken within the virtual
world (e.g., simulating reinforcement, enemy counter-attack, etc)
will be relayed to the ground troops, which may then be taken into account in the ground decisions. For example, if a region in the
ground occupied by troops were air-raided, then the troops should
``perish''.
Figure~\ref{fig:military} illustrates possible scenarios of employing metaverse in military training and modern wars.
Unfortunately, with the advancement of satellite systems, unmanned aerial vehicles, supersonic drones and defence technology,  modern warfares involving intensive cyber-physical simulation and deployment are likely to be more destructive. 

\begin{figure}[h]
\centering
\includegraphics[width=2.8in]{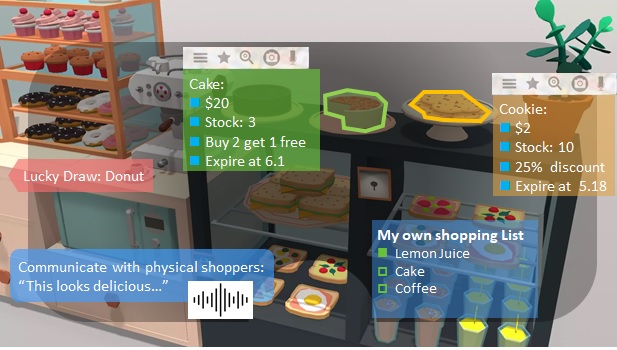}
\caption{Market Place}
\label{fig:marketplace}
\end{figure}

\subsection*{The Marketplace}
In today's marketplace, we either shop in a mall or 
buy our products online. A physical shop may have
a web page (a very primitive form of the virtual model), but there is
very limited real-time interaction between the two spaces. 

In tomorrow's metaverse shop,
a physical space may be ``expanded'' to display more items.
Figure~\ref{fig:marketplace} illustrates a co-space shopping scenario in a metaverse pastry shop.
Similarly, in tomorrow's metaverse marketplace, a physical mall will be
``expanded'' into a mall (virtually) that houses many more shops
than the physical mall. In addition to the virtual correspondence
of the physical shops, the virtual mall can rent out virtual space
for virtual shop owners. In the physical mall, screens (large
displays) can be set up within each physical shop for cyber
shoppers to communicate with physical shoppers within the same
shop (e.g., through simple text messages and/or more sophisticated 
immersive technologies). While a physical shopper is
restricted to the shops that are physically located in the mall, 
the online shopper has a wider selection of shops (and products).
The virtual mall needs to be kept up-to-date with real-time
information from the physical mall, e.g., live programs that are
happening in the physical mall, ongoing lucky draws, updates on
the availability of products, etc.  
The same concept may be used to ``expand'' the space
for exhibits in a museum, or even ``expand'' a stadium's sitting capacity for the global audience: sports enthusiasts may view a game through the ``eyes'' of the player, and concert audiences may have close-up immersive experience.


\begin{figure}[h]
\centering
\includegraphics[width=2.8in]{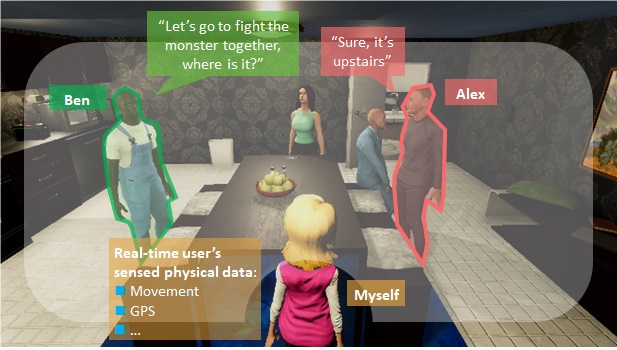}
\caption{Gaming and Social Networking}
\label{fig:gaming}
\end{figure}

\subsection*{Gaming, Social Networking and E-commerce}

One class of gaming in the metaverse environment is location-based gaming
(LBG). LBG is gaining popularity and is believed to
be the future of video gaming where a player's everyday experience
(e.g., of moving around the city) is interleaved with the extraordinary
experience of a game.
These games deliver an experience that
changes according to the player's locations and actions.

In LBG, a user equipped with a GPS-enabled handset (e.g., a mobile
phone) can play a video game that combines a player's real world
(i.e., his physical location) with a virtual world on the handset.
The physical location becomes part of the game board,
and the player's movement directly influences the gaming progress
(and may affect the game character and/or environment). 
Pokémon GO \cite{pokemon} and
{\it Zombies, Run!} \cite{zombies} are examples of LBG.

Another form of metaverse games integrates a physical environment
with a corresponding virtual model. Here, RFIDs and sensors are used
to capture the information about the players' current context, which
is transmitted to a server. The server (which may be controlled by
another player) follows the game rules and
relays back to the physical players some information that helps them to
proceed (e.g., locations of enemies in the vicinity).
Examples of this category of games include
Tourality \cite{tour} and Wanderer \cite{wanderer}.

Social networking can also be conducted in the metaverse. A person in a
certain location in the physical space can detect a
friend at the same location in the virtual space, and they may 
fight together against some enemies in the virtual space.
Similarly, two ``comrades'' who
fight together in the virtual space can detect each
other when they are near to each other in the physical space,
giving opportunity for more social interaction.
Users may form interest groups to share
information and artifacts, and trade user-created contents and virtual valuables, including non-fungible tokens (NFT). 
They may use the metaverse to conduct consumer-to-consumer or business-to-consumer sales of physical world products.
Figure~\ref{fig:gaming} illustrates such an example.

It would be interesting to see how the multi-billion dollar
industry of e-commerce, games, and social networking will grow as advances in
technologies to support the metaverse become mature.


\begin{figure}[h]
\centering
\includegraphics[width=2.8in]{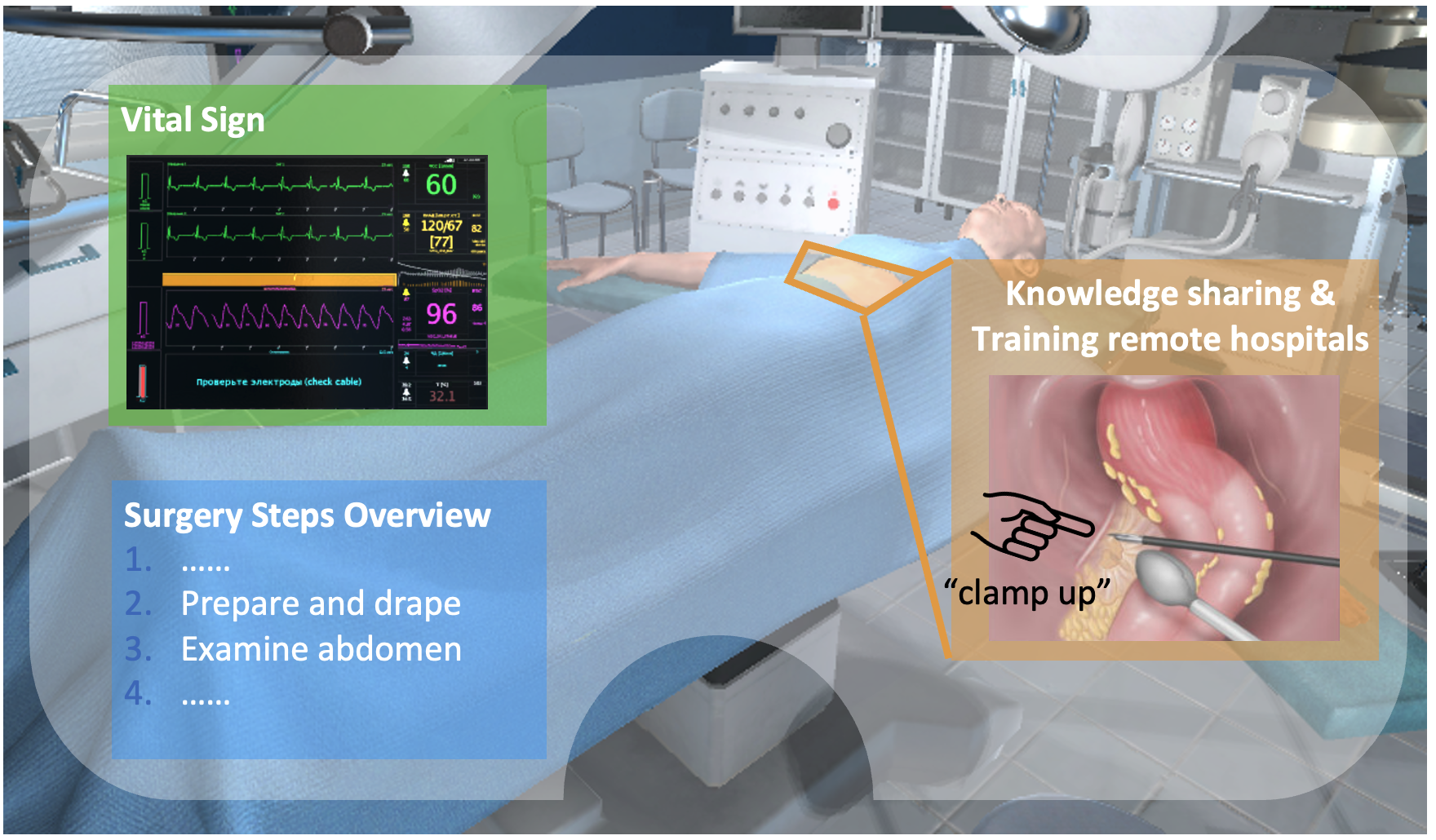}
\caption{Assisted Surgery}
\label{fig:surgery}
\end{figure}

\subsection*{Smart Healthcare}

Healthcare has been improving over the years, with the use of AI, sensors, and improved hardware and network bandwidth, etc.  
With the onslaught of the pandemic, Covid-19, we have seen increasing acceptance of telemedicine,  remote consultation, and online collaboration.
Telemedicine, or part of the examination process if not totally,  has improved the efficiency and comfort of patients.
Clinicians are able to monitor and diagnose many chronic cases or minor illnesses or understand the conditions efficiently.

With the metaverse, virtual healthcare could be improved with virtual reality (VR).
In fact, holography-guided heart surgery is already used at National University Hospital to enable more precise and speedier incisions~\cite{nuhcs}.
As illustrated in Figure~\ref{fig:surgery},
remote assistive surgery can become a real possibility with VR, thereby improving knowledge sharing and training in remote hospitals and clinics.
Psychologists and psychiatrists can use VR in aversion therapy to help resolve anxiety issues or overcome
delusions. It may be used to
replay the situation to overcome the fear or guilt, and treat post-traumatic stress disorder (PTSD).
Metaverse enables treatment in private and an environment where interactions can be captured for analysis.

Radiology enables the clinicians to pinpoint the critical spot of the affected organ that requires medical attention or treatment.
Its scanning technology,
augmented with AI technology,
has made great advancements in terms of quality and precision, which has resulted in more accurate diagnosis and better healthcare outcome~\cite{radiology21}.  
As ``side effects'',  early detection of some diseases such as cataracts may be made possible through image capturing with the headset and AI-enabled analysis of eye images, without bulky and expensive devices and going to the hospital.
Metaverse, with more advanced visualization and real-time comorbidity analysis, may enable surgeons to make better decisions in surgery and treatment.

\subsection*{Smart City}

Smart city refers to a technologically modern urban area that makes use of technology for technological-aided  living and city planning and management.
Sensors, cameras, VR/AR and fast network connectivity are needed to enable such a living condition. In metaverse, interactive platforms support urban environments modeling in 3D and with huge amounts of point cloud and infrastructure data, thereby enabling better city planning for efficiency and comfort.
This is somewhat related to the concept of Digital Twin.  However, metaverse, to a great extent, has ``life'' of its own, with data being created and exchanged by human beings in the physical world.

Metaverse not only improves the quality of living, making life a lot more convenient and richer, but also, enables efficient management of various resources which helps reduce carbon footprints 
based on data-driven decision making in tweaking our infrastructure to account for changes in human behavior.
In a smart city, tourists may retrieve information with visualization, and preview the daily events on offer before making decisions.
Likewise, people can choose to work from home, and the morning commute may become a thing of the past.
In summary, smart healthcare, marketplace,
transportation,
environmental, social and governance (ESG) implementation, together with advanced infrastructure, form the fabric of a smart city.

\section{Observations from Data-centric Perspectives}
\label{sec:challenges}

From the above discussions, we have the following observations of the data-rich metaverse:
\begin{itemize}
\item
There is a large amount of data/information
generated within the metaverse.
Some of these are static (e.g., maps, quantity-on-hand),
while others are dynamic (e.g., locations, sensor data) and frequently
changing. Moreover, the data may come in different formats (e.g., non-structured
like video and textual and structured like personal data). These data may
also come from multiple different data sources (e.g., static data from
a relational database, video data from a video streaming service).
In addition, the large amount of data may have to be streamed from
one space to another, particularly from the physical to the virtual
to ensure real-time tracking of the environment.
To a great extent, the data, generated and collected in real time, may break the 3Vs (volume, variety, and velocity) properties of the Big Data.
\item
There is a large number of sensors that are
used to capture the data from the physical environment.
In-network processing may be needed to aggregate data before
transmission.
\item
There is a large number of events generated within the metaverse.
These have to be monitored, and
may trigger further actions/events both in the physical and virtual worlds.
\item
There is a large number of users (and queries). Each user device basically
contributes a distributed node into a highly distributed environment.
\end{itemize}

%
Clearly, the database community has been dealing with the above-mentioned
(perhaps, not at the scale that the metaverse entails). 
We have addressed and are addressing a wide range of research
problems~\cite{seattlereportcacm22} that are relevant - sensor networks, data streams,
distributed databases, update-intensive operations, database sharding, workload partitioning, data and system security,
search
and data retrieval, and so forth. 
As such, 
we should be able to
contribute to this new field and chart the research directions
ahead.


\section{The Metaverse Challenges}
\label{sec:detailchallenges}
Being an integration of the physical and virtual spaces,
it is certain that the metaverse brings with it the research issues
within each space. In the physical domain, we need to design
efficient and effective methods to sense the physical environment
(through extensive use of RFIDs or sensing devices), to
transform these data into a form that users will appreciate
(through data cleansing, data mining, aggregation or interpolation), and
to process queries in-network, and so on. While
some work have been done (e.g., \cite{Gonz07,Jeff08,Madd05,Welb07, wang2022, DBLP:journals/corr/abs-2202-10336}),
we are only scratching the surface to realize practical deployment.

In the virtual space, with the popularity of Massively Multiplayer Online
Games, there has been a tremendous amount of interests in recent
years to design techniques to support interactive virtual environments
for users to communicate with each other in real-time
\cite{Gupt08,Whit07b,Whit07,Zhou04}. As pointed out in \cite{Whit07}, there are
a number of research challenges that need attention, including
designing database engines for games workloads and methods to guarantee
consistency across multiple virtual views. Techniques for caching
and indexing virtual environments (e.g., \cite{Shou01,Shou04}) need
further study to scale to the large number of users.

For the rest of this section, we shall focus on challenges that
arise as a result of the integration between the two spaces that
may be of interests to the database community. Some non-database
related issues, which we have left out, include (a) novel interface technologies that can
seamlessly link the physical and cyber spaces to support real-time
interaction between users within the two spaces; (b) innovative
visualization and presentation of output (events and data)
within the metaverse on a wide range of devices and platforms
(small vs large displays, fixed vs mobile); (c) techniques, tools
and devices for capturing data from the physical environment (input methods fusing voice and gesture via a head- and hand-piece in place), and
for creating content (high-fidelity digitization of humans, scenes and interactions, 
animation and effects, immersive experience) for the virtual environment; (d) language
translation, transcription and mediation methods to support
social networking and learning, and many others (e.g., security and
networking infrastructure).
With more users and activities, and a bigger surface area for attacks in the metaverse,
new kinds of cybercrime and frauds are likely to emerge, and these definitely need to be addressed in the future. 

\subsection{Data Fusion over Heterogeneous Data \\ Sources}
Data fusion is generally defined as the use of techniques that
combine data from multiple sources through inferences in order to
produce data that is potentially more accurate than if they were
obtained from a single source \cite{Hall04}. While data fusion
has been studied in the context of sensor networks, 
data fusion in the metaverse is more challenging as the inputs may
come from a wide variety of sources including 
blogs, video/audio clips, and photographs about
events that took place in the digital and physical worlds.

\begin{figure}
\centering \epsfig{file=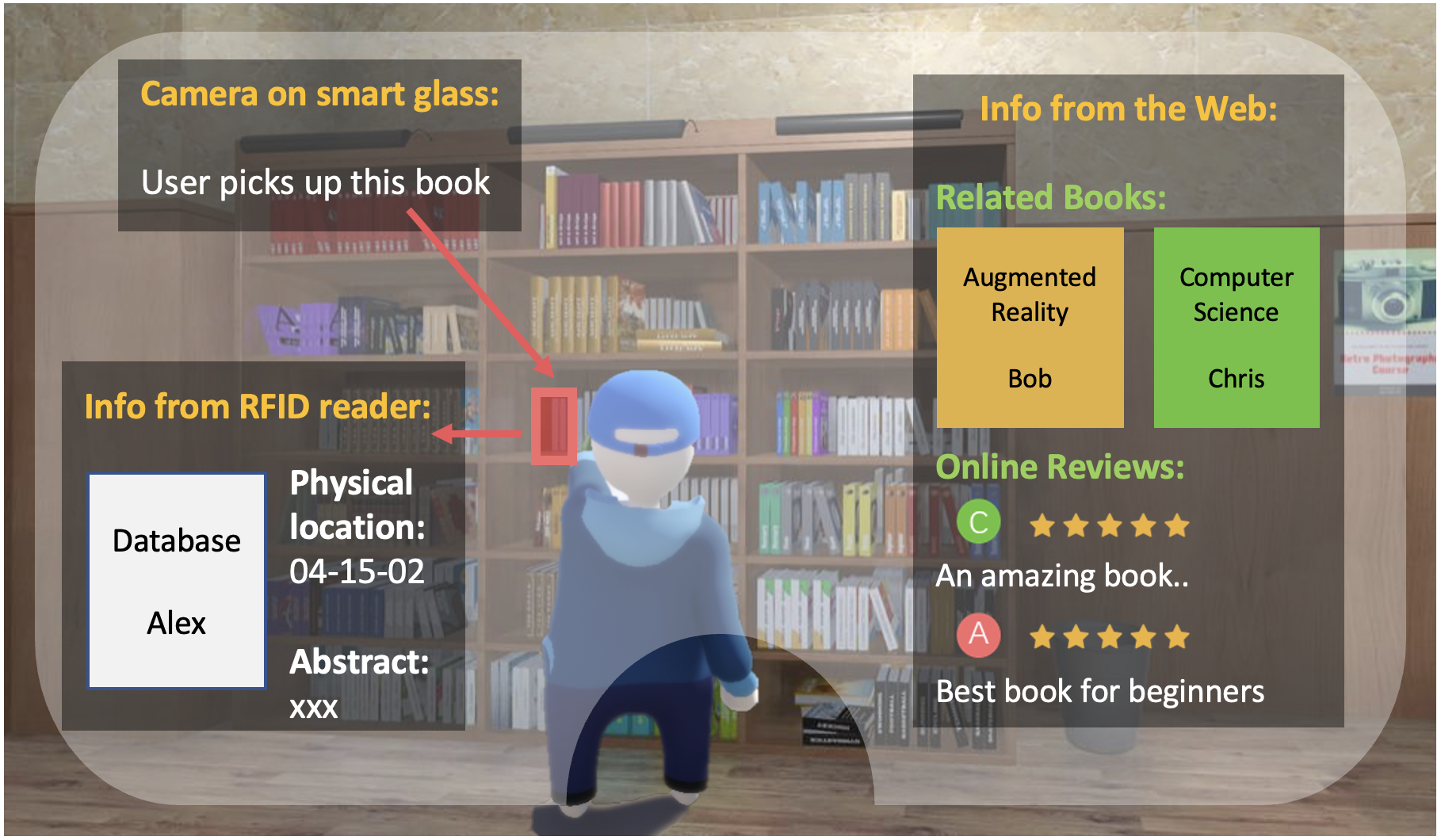,width=2.8in} \caption{The
Co-space of a Library} \label{fig:lib}
\end{figure}

As an example, consider the metaverse of a library in Figure
\ref{fig:lib}, information from both video camera and RFID readers
will be needed to ensure that the location of books are
represented accurately in the digital space. Furthermore, reviews
and opinions on the books can also be drawn from both the Web and
the social networks of the users to enhance their browsing experience.
Such fusion of information on a single entity requires a
substantial amount of inference over semantics that are extracted
from multiple data sources.

From the above discussion, we note that the metaverse data management is related to
the well studied fields of data stream processing
\cite{ZhouAST08}, sensors network \cite{Madd05} and data
integration \cite{Lev00,Len02}. However, it also differs in at least two
ways. First, unlike the relatively simple aggregation that is
being done over data streams and sensors presently, managing data
in the metaverse requires more complex logic inferences over these data
sources. Second, unlike data integration which aims to derive a
common schema for a set of heterogeneous databases, the metaverse data
management detects events that had taken place based on these data sources and
depicts these events accurately and efficiently in the
metaverse.

There is a clear need to develop data fusion mechanisms that
can deal with these two issues effectively. Recent works on
polyglot data management offer a good starting point here
\cite{poly1, poly2, poly3}. 

\subsection{Data Collaboration}

Data collaboration should be  more prevailing in the metaverse, as multiple parties may pool data for more meaningful data insights.
However, data misuse and unlawful data
collection have led to precautionary measures being imposed by
individual organizations to guide against data leakage and abuse.
Privacy-preserving data and knowledge sharing mechanisms with fair contributions of useful data have to be designed.
To promote data collaboration 
and to discourage  free-riders from intentionally obtaining the others' data and parameters without doing their part, effective and computationally efficient incentive models have to be designed. 
In the metaverse, the users are likely to be heterogeneous in data qualities and quantities, possibly with non-independently and identically distribution (Non-IID), and computational powers.  
Such heterogeneities complicate the design space of collaboration strategies and mechanisms.



\subsection{Data Consistency}
In networked virtual environments, it is important that users have
a consistent view of the virtual world. This requires transmitting
data within the virtual world. Unfortunately, there is to-date no
solutions that can scale well. Now, in the metaverse, the requirement
of consistency becomes even more challenging - the virtual world
must also reflect what is happening in the real world. Given the
constraints in bandwidth and the large amount of data to be
transmitted, we do not expect to see a truly consistent view in
both worlds. However, we can try to keep the virtual world as
close to the real world as possible. One solution is to tolerate
some degree of discrepancies - for numerical data, they may be
within certain coherency requirements; for multimedia data, a low
resolution image/video may be used instead. Some recent works have
looked at how to disseminate streaming data to a large number of
clients while preserving data coherency
\cite{Bhid07,Shah03,Zhou08}. These techniques assume a small
number of distinct objects, and so do not scale to a large number of
objects.

A closely related approach is to study how data to be
transmitted should be prioritized.
For example, more critical data
can be transmitted first before less critical data.
We can learn from methods developed for intermittently-connected
and disruptive networks \cite{Zhan07}.
We believe there is much needed avenue to be explored in this aspect,
e.g., to study different scheduling schemes. Besides 
prioritizing
data, it may also be necessary to develop techniques to
schedule multiple (continuous) queries that meet different
Quality of Service (QoS) metrics. While techniques developed in 
\cite{tods08} provided some insights on how this can be
effectively handled, we believe this direction deserves further
investigation.



\subsection{Security and Data Privacy}

In the metaverse,
the increasingly interconnected and complex systems and networks increase our exposure 
to various threats, such as the loss of data, the loss of control over the software systems 
and networks, the mistrust due to ill-intended contents and actions, and so on.  
As a consequence, data security and privacy are major concerns.

To enhance data security in the metaverse, there exist various emerging technologies that could enable trusted collaborations and interactions, through secure communication, transfer and exchange of digital documents and assets, verifiability, and auditability. For example, commercial secure hardware can be used to establish trusted execution environments (TEE) where data and computation 
cannot be modified by any other parties, including cloud vendors or service providers; then, encryption techniques can be used to protect the data and models in communication channels and on untrusted storage on the cloud. 
However, despite the promise of a trusted
execution environment, current implementations like Intel SGX fall short of security (vulnerable to
side channel attacks) and performance (large overhead). 
As another example, Blockchains~\cite{ruan2022blockchains} can serve as the basis for connectivity in the metaverse to make it open and decentralized. Transactions among different parties in the metaverse can be permanently recorded and verifiable in blockchains, thus fostering a wider acceptance of the metaverse. These new applications require a re-design of data management and processing systems. For instance, decentralization requires the computation to be byzantine faulty tolerance, which introduces a huge cost in replication and consensus modeling. One possible solution is to use verifiable ledger database systems~\cite{ledgerbench2022, glassdb} with a trusted third party serving as the auditor.
%
To  design an efficient verifiable ledger database.
we should consider three factors:
abstraction, threat model, and
performance. 
The system may combine
efficient cryptographic techniques, often found in authenticated data structures such as the Merkle Tree, and transparency logs, with trusted
hardware, and distributed ledgers. 
Such a system must support distributed transactions and have efficient proof
sizes, and be flexible, so that it can
be configured to achieve different trade-offs in the design space to fit specific use. 

Meanwhile, protecting data privacy in the metaverse requires a delicate balance between minimizing privacy risk and maximizing data utility~\cite{wang2022}. On one hand, mitigating privacy risks requires restricting access to sensitive data and conforming to the privacy regulations introduced by the governments. On the other hand, data-centric applications in the metaverse may rely on the availability of various types of data pertinent to users. To address this tension between data privacy and utility, it is important to develop new algorithms and paradigms to enable data analysis in a privacy-preserving manner. Towards this end, there are emerging technologies such as federated learning~\cite{KairouzMABBBBCC21,xxk2021,wuvldb202} and differential privacy~\cite{DworkMNS06}, but major research effort is still needed to make them applicable for the metaverse applications. In addition, incentivizing users to share information for privacy-preserving analytics is a challenging issue~\cite{luoicde2021}.




\subsection{Database System Architecture}

With a large number of cyber users, and physical users with
handheld devices, the metaverse platform naturally forms a
highly distributed (peer-to-peer) system. The system is highly complex
because of the heterogeneity of the devices. Moreover, there is an
enormity of static and dynamic data that flow within each space
and across spaces.

For queries that access static data that are stored locally,
techniques that can facilitate search/discovery of relevant
information are critical. P2P search methods may be applicable
here \cite{Hueb05,Jaga06,Wu08}. However, for dynamic data that
need to be streamed from one space to the other, these methods may
not be suitable. While there has been considerable work on
distributed stream processing \cite{Abad05,Chan03,Hwan08}, these
are restricted to query processing and typically assume a smaller
number of sites and do not address the heterogeneity across the
sites. Here, it seems that publish/subscribe architecture
\cite{Eugs03,Gupta04,Zhou08,ps15,pubsub20} may be more effective.
Novel architectures that can support streaming data and search
efficiently are needed. For example, we envision a
publish/subscribe system over peer-to-peer networks where each
peer may be a highly parallel cluster that can support a large
number of mobile clients.

The need for supporting a large number of concurrent and
both data and computational intensive activities, requires
new system architectures to be autonomic and adaptive and
scalable. 
In addition, the loads need to be adaptively balanced and new
nodes can be easily added without substantial reconfiguration effort.
As an example, let us consider product promotions or flash sales like ``Black Friday'' in metaverse shops.
Compared with conventional databases for e-shops, the metaverse databases should scale more efficiently.
Because during the sales, metaverse databases need to handle large amounts of requests not only from the virtual shop, but also from the physical shop.
%
Hence, managing the metaverse data calls for a re-examination of the database architectures as we understand today.

Recently, the processing paradigm of MapReduce \cite{mr} and 
other similar applicative programming frameworks have 
revolutionalized the extreme data analysis on
clusters, and systems such as Clustera \cite{clustera} exploit modern
software building blocks for efficiency and scalability. 
These
and some other recent efforts in exploiting multi core architectures and commodity hardware (e.g., GPUs, FPGAs, non-volatile memories) and hardware/software co-design
may provide a basis for development of new database engines~\cite{DBLP:journals/tkde/ZhangCOTZ15, DBLP:journals/sigmod/TanCOWYZ15}. 
Efficient resource allocations and utilization~\cite{metaslice2022}, and new processing paradigms such as serverless computing~\cite{wu2022serverless} are important for  efficiency and scalability.  

More recently, there has been works done to develop novel architectures
for managing visual data. Most of these works manipulate the visual 
data as video data. As an example, LightDB was proposed to handle
augmented, virtual reality and mixed reality data in the form of
video content \cite{lightdb18}. While such approaches are promising,
they are restricted to applications that are static. 
%
%
In contrast,
we are
dealing with (a) a large amount of diverse types of data, ranging
from structured to unstructured, textual to video, static and dynamic;
(b) data that exist in two different spaces;
(c) with a large number of cyber users, and physical users with handheld devices, the database has to ensure better efficiency, flexibility, resiliency, and elasticity than it
has today for the metaverse.

We now discuss design considerations for metaverse database architectures by 
examining the
emerging areas such as
decentralization, disaggregation, and serverless.

\subsubsection{Decentralized}
%
%
The data volume today is far more than a single database instance can handle.
Hence, decentralized databases, storing data across a network of distributed servers, possibly owned by autonomous individuals, become increasingly popular due to the movement on Web3.
Some application specific metaverse databases, that demand high degree of transparency,  are expected to explore peer-to-peer based and decentralized architecture for highly scalable data services.
In the metaverse, users are increasingly interconnected, no matter within a region, across countries, or even across continents.
We envision that metaverse databases would be worldwide decentralized.

Due to the decentralization, distributed transactions are essential for accessing data across multiple data centers.
However, distributed transactions are hard to process at scale to ensure
high throughput, high availability and yet low latency 
due to the network partition and non-negligible inter-data-center network latency.
Although existing works~\cite{DBLP:conf/eurosys/KraskaPFMF13,DBLP:conf/sigmod/YanYZLWSB18} on reducing network overhead for inter-data-center transactions can potentially help, 
horizontal scalability in decentralized metaverse databases remains a challenging issue.
This may consequently further spawn new research areas, such as
design of high-speed networks, efficient database sharding and high availability, and decentralized transaction protocols.

\subsubsection{Disaggregated}

\begin{figure}
  \centering
  \includegraphics[width=0.44\textwidth]{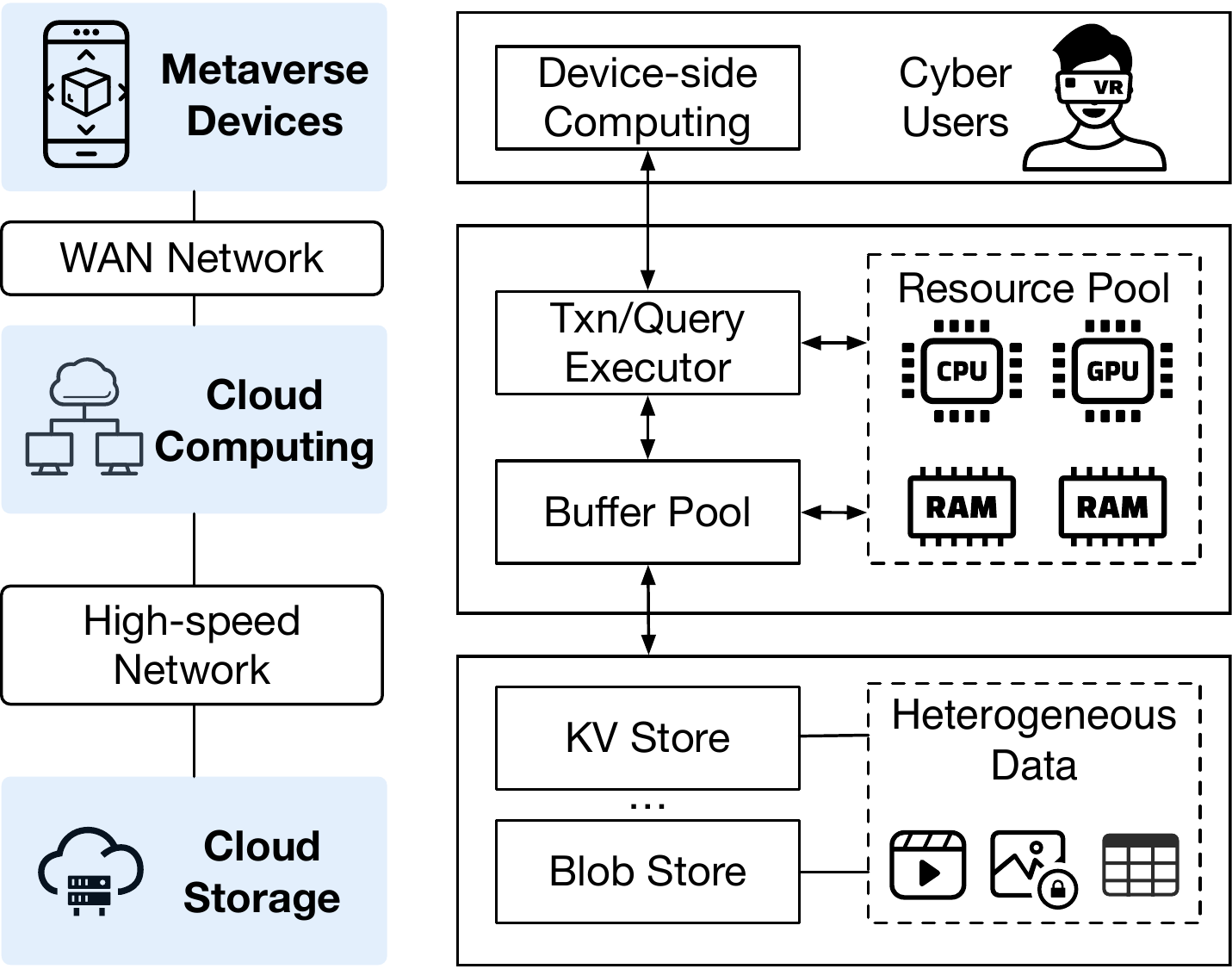}
  \caption{Disaggregated Architecture of a Metaverse Database}
  \label{ovi:disaggregated}
\end{figure}
In recent years, with the evolution of cloud infrastructures, cloud-native databases have become mainstream.
The foremost innovation of cloud-native databases is to disaggregate the computation and storage into two distinct components, enabling each to scale independently.
Such disaggregation facilitates efficiency and elasticity, which are pivotal for managing metaverse data.
Following this trend, metaverse databases can be designed as cloud-native with disaggregation.
Besides, we observe that device-side computation is not considered in today's disaggregation architecture.
Unlike traditional users, cyber users enter into and fully immerse themselves in the metaverse world through specific metaverse devices, such as virtual reality goggles.
These devices are equipped with increasingly powerful processors to collaborate with cloud metaverse databases, enabling part of the computation to be further separated from the cloud side to the device side.
Hence, we envision a new disaggregation architecture, namely device-cloud-storage disaggregation, for tomorrow's metaverse databases.



Figure~\ref{ovi:disaggregated} illustrates a design of device-cloud-storage architecture for metaverse databases, which consists of the following three components.

\noindent
\emph{Metaverse devices.} Cyber users access data in the metaverse using metaverse devices.
By exploiting the equipped large processors, these devices can afford part of computation tasks like data aggregation and fusion.
Some works on processing in portable devices \cite{Kaln06,Mamo03} and energy-efficient optimization \cite{Alon93} can be potentially extended for device-side computing.

\noindent
\emph{Cloud computing.}
This layer is responsible for query processing, transaction management, etc.
In this layer, the \emph{transaction/query executors} handle the incoming transactions/queries sent from metaverse devices, and fetch data from the \emph{buffer pools}.
These transaction/query executors and buffer pools can scale elastically based on the workload,
by leveraging on-demand resources from the resource pool.
With the explosive growth of cloud infrastructures, 
computing and memory resources are likely to increase further and should be able to efficiently support metaverse databases.
To reduce network overhead for transferring the data from the storage layer to the buffer pools,
we can cache data in the buffer pool as much as possible, and
use high-speed networks such as InfiniBand and remote direct memory access (RDMA)~\cite{rdma2018}.

\noindent
\emph{Cloud storage.}
This layer hosts the data and logs produced from the virtual and physical spaces. 
To efficiently handle a large amount of diverse types of data, ranging from structured to unstructured and textual to video, this layer contains heterogeneous data stores, including the key-value (KV) store, object store, block store, etc.
This layer can be constructed by employing various cloud storage services that ensure scalable, high-performance, and high-available storage.
For example,
in Microsoft Azure Cloud,
it is possible to
deploy the KV store on Azure cloud Redis service, while using Azure blob storage and Azure disk storage respectively for the object store and block store.


\subsubsection{Serverless}

Serverless computing~\cite{hellerstein19, berkeley-serverless} has emerged in response to the rise in popularity of cloud architecture.
It enables application developers or owners
to focus on the application logic and leave the infrastructure-level tasks, such as resource management and auto-scaling,
to the cloud vendor.
It is therefore an efficient method to scale microservices as well as monolithic applications in the cloud, 
in line with the principle of
workload based
resource allocation. 

Serverless computing comes with two key requirements: (a) \textit{ease of management}; and (b) \textit{fine-grained pricing model}.
First,
from the clients' perspective, they only need to upload the execution logic and define the trigger upon which the job is executed, e.g., an HTTP URL~\cite{serverless-in-the-wild}. 
The cloud vendor takes care of creating and revoking serverless instances as well as scheduling the requests.
Second, clients are charged based on the actual amount of resources consumed during execution, with fine-grained granularity similar in spirit to pay-as-you-go charging model.
While the benefits
of serverless model serving are well acknowleged~\cite{wu2022serverless}, 
the security limitation, that requires the cloud to be trusted, remains.
Indeed, the cloud can be compromised, either by malicious insiders or by exploiting software vulnerabilities, and once compromised, the entire software stack, including the operating system (OS) and hypervisor, will be under the adversary’s control.
For security, TEE has been exploited, by partitioning the
application logic 
into a trusted part, 
which runs inside the TEE enclave, and an untrusted part that interacts with the OS.
While the new generation of TEE has increased memory capacity, the code base still need to be optimized for efficiency and reducing frequent reloading.
There remain many efficiency and security problems to address in this design space.

\subsection{Organization of Data}
While it is clear that data of different types need to be managed
separately, it is not immediately clear that data of the same type
from the two spaces should be treated separately. In other words,
should the location of a shopper in the physical mall be stored
together with the location of an online shopper; or should the
real-live images of a museum be handled in the same way as the
virtual images. On one hand, we can simply tag data to reflect the
space it belongs to. This offers a unified view of the metaverse
and simplifies the management of data. However, for operations
that involve only data from a particular space, the performance
may be penalized. On the other hand, we can organize the data from
the two spaces separately. But, this may end up duplicating
resources. Moreover, it may be possible to have a hybrid strategy
- for certain data types, integrating them may be the best; for
others, keeping them distinct may be optimal. 



The metaverse would have a huge amount of trajectory and virtual walkthrough data, and to facilitate efficient retrieval, efficient indexes are needed.
For example, in
\cite{Shou04}, an HDoV tree was proposed to index content at
different degrees of visibility in a virtual walkthrough
environment. This structure is obtained statically, and requires
high computation overhead.
In the metaverse, we may need a more robust
and dynamic structure to cater to the frequent updates of
information. While some work has been done for locational data
\cite{Chen08,Jens04}, no such indexing methods have been designed
for the virtual domain. We need more flexible schemes, in-memory-, solid-state storage- and disk-based,  to
handle update intensive applications and frequently changing
scenes.


The two categories of data (coming from the physical and virtual
domains) also call for novel buffer management and caching schemes.
In particular, we expect an effective scheme to be conscious of the
semantics. 
For example, data from the real space may be given
higher priority over data from the virtual space. However, we need
to develop criteria to compare
the priorities across the two domains.

\subsection{Query Processing and Optimization}

Query processing and optimization in the metaverse will
require novel mechanisms. First, new operators may have to be introduced.
As an example, sensor data may have to be interpolated (or combined
using some user-defined functions) for them to be
consumed by the virtual space. In fact,
data can be processed and transformed as soon as they are received;
alternatively, they can be transformed at runtime.
As another example, data in the virtual
space may be interpreted in a different way from those in the physical
space. These inevitably led to changes to the optimizer to become
aware of these operators in order to generate an optimal plan.
Earlier work~\cite{Hell98} on optimizing queries with expensive
predicates may offer a good starting point.

Second, the performance requirements for the two spaces may not
be necessarily the same. For example, it is reasonable to prioritize
sales for a shopper in a physical mall than for an online shopper
(when they both wanted the last available item).
As another example, in the case of a cyber user, while real-time
information is highly desirable, approximate data may be tolerated
(e.g., instead of a high resolution video stream, a low-resolution
stream or animation may be acceptable). This calls for query processing
or optimization techniques to be ``space'' aware. Moreover, efficient
approximation techniques in the virtual space that do not sacrifice the
quality of the output significantly are highly desirable.

Third, besides I/O, CPU and bandwidth considerations, the optimizer may
have to be device-aware so that a feasible (and optimal for the device)
plan can be generated. Some works on processing in portable devices
\cite{Kaln06,Mamo03} and energy-efficient optimization \cite{Alon93}
can potentially be extended for the metaverse.

Fourth, we are dealing not only with moving objects (some
moving in the physical space), we are also dealing with
moving queries (a user moving in the virtual environment may need
to track all users within his/her views - as he/she moves, his/her views
of the space changes). There are very few works on moving queries
over moving objects \cite{Gedi06a,Gedi06b}, and this area is certainly
worth further exploration.

Fifth, one key challenge in designing a distributed architecture is
to ensure that meta-data that are required for optimization
can be estimated locally at each site/cluster to minimize
information exchange, while at the same time the quality of the
generated plan may not be significantly compromised.
Designing such a cooperative system is difficult.
Techniques from distributed databases may be relevant here
\cite{Ston96}.

Finally, the metaverse produces huge amounts of data, in the form of data streams. Stream and in-situ or schema-on-read processing become important.
To sustain high stream ingress traffic, data processing
operators have to be replicated and run
in parallel threads.
Stream execution plan optimization and exploitation of appropriate and new hardware are necessary
~\cite{DBLP:conf/sigmod/ZhangHZH19}.
Parallelism for scalability may be achieved by optimizing the processing on the cores and servers \cite{highspeed2015, hebsstream2021}, and efficient remote processing based on remote direct memory access (RDMA)~\cite{rdma2018}.
We shall discuss this more from the perspective of AR/VR in Section~\ref{ARVR}.

\subsection{Intelligent Decision Making and Self Driving Optimizations}

In recent years, we are witnessing a significant amount of effort to integrate
database systems and machine learning (ML). On one hand, with ML being more data-centric, 
there is a need for database support to manage 
the data efficiently and effectively. On the other hand, 
incorporating learning into database systems  
facilitates the design of self-tuning, self-healing, and autonomous DBMS.
Given the applications in the metaverse are as complex, if not more, as 
modern applications, data-centric ML is required for 
effective predictions and recommendations to support informed decision making.
Moreover, to ensure efficient processing, novel ML mechanisms will need to
take advantage of the rich data in the metaverse. 
Regularization, feature interaction, explanation, efficient communication, and many other optimizations can be viewed and addressed from a data-centric perspective
~\cite{DBLP:journals/sigmod/0059Z0JOT16,
DBLP:conf/aaai/LuoCCO021, wudml2021}.

This trend of DBxAI is set to continue for better integration and exploitation 
of the two technologies~\cite{armnet, RLforDB2022}. However, to really make the data system 
scalable and intelligent, 
we may need to re-examine the needs and requirements 
for making AI/ML an integral part of the system, instead of putting
an AI/ML layer on top of existing systems.
In other words, learning from a particular instance of dataset and 
query patterns may only improve database optimization techniques, such as cost estimation and index design, and their system performance only temporarily.
The fact that databases are dynamic in nature may make the AI/ML models and algorithms ineffective due to data and feature drift problems. This is more so in the metaverse, where the users and applications are more diversified and dynamic. From the perspective of AI and machine learning, a recent survey explores the role of AI in the foundation and development of the metaverse~\cite{DBLP:journals/corr/abs-2202-10336}. 

\begin{figure*}[t]
    \centering
    \begin{minipage}{\linewidth}
        \centering
      
        \subfigure[Conventional Learning]{
            \centering
            \includegraphics[height=4.5cm]{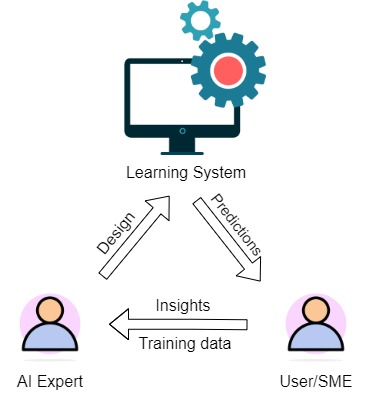}
            \label{fig:int1}
                   }
                   \hspace{3mm}
        \subfigure[Self Learning]{
            \centering
            \includegraphics[height=4.5cm]{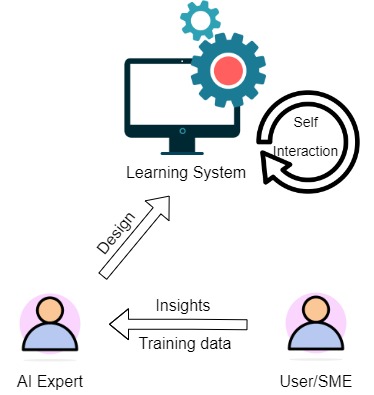}
            \label{fig:int2}
        }
        \hspace{3mm}
        \subfigure[Interactive Learning]{
            \centering
            \includegraphics[height=4.5cm]{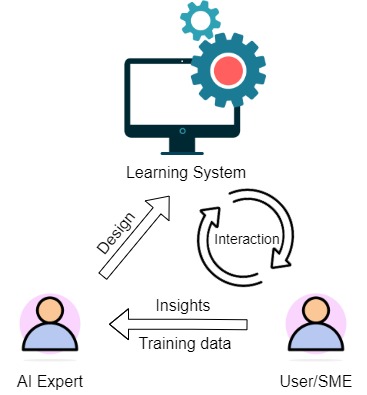}
            \label{fig:int3}
        }
        \caption{Learning Workflows}
        \label{fig:learningworkflows}
   \end{minipage}
\end{figure*}

\subsection{AR/VR Data Streaming and Learning}
\label{ARVR}

To build up the metaverse, an integral step is to digitize humans, objects, and scenes from the physical space into the virtual space. 
Nowadays, many AR/VR applications either only (1) have low-fidelity digital avatars and scenes that 
we can easily tell they are not real, or (2) still heavily rely on artists to manually create the digital counterparts, which is time-consuming and not feasible for general users to adopt. 
Recently, a new digital twin technology, i.e. Neural Radiance Fields (NeRF) \cite{mildenhall2020nerf}, demonstrates unprecedented high-fidelity in automatically reconstructing human-like avatar \cite{tsutsui2022novel}, realistic objects \cite{liu2022devrf} and large scenes like a city \cite{tancik2022blocknerf} in the metaverse.
A metaverse of high-fidelity could provide users immerse experience and unlock some new applications such as watching a basketball game from any viewpoint.
Despite being promising, a key challenge towards high-fidelity is data explosion. This is because, in order to look realistic, avatars need to exhibit skin-level details and dynamic objects need to exhibit accurate movements, let alone much larger scenes like cities. In contrast to learning a representation for each avatar or object independently, a promising research direction is to create generalizable representation that can be shared among similar avatars or objects, and develop algorithms to efficiently customise, store, and operate the digital assets given the new representations.

Once the metaverse has been built up, the large amount of data generated in the metaverse offers great opportunities for developing AI models specifically for AR/VR. 
In particular, different metaverse users may have different tasks that need AI to assist, making it challenging to customize AI model and training data for each use case.
A promising solution is to develop foundation deep learning models \cite{bommasani2021opportunities,brown2020language,wang2022,cai2022revitalize} that are first trained on large-scale data and then can be applied to various downstream tasks with minimal task-specific fine-tuning. 
As a starting point, the recent work EgoVLP \cite{lin2022egocentric} conducted visual-language pretraining on Ego4D \cite{grauman2021ego4d}, a massive egocentric video dataset that targets collecting what smart glasses see throughout human daily activities, and then the pretrained EgoVLP model demonstrates strong performances on various metaverse applications such as action recognition, moment query via a textual description, object state change detection. 
Due to the large data scale, pretraining often requires considerable compute resources and takes a long time. For example, pretraining in EgoVLP \cite{lin2022egocentric} takes two days on 32 A100 GPUs. 
This is a major roadblock for developing more advanced pretrained models, e.g. continual learning and update of the pretrained model given the newly arrived data stream \cite{wu2022label} in the ever-changing metaverse. 
To overcome this challenge, it is important to optimize data communication between storage and compute nodes, and to exploit parameter-efficient continual learning and transfer learning of large deep learning models.

The conventional workflow of machine learning has limited human-machine interaction. As Figure~\ref{fig:learningworkflows} (a) illustrates, it is mostly unidirectional as only the machine learns from humans on “how to work”. 
Figure~\ref{fig:learningworkflows} (b) outlines another popular machine learning workflow i.e. self-interactive machine learning, which is used in AlphaGo where an AI agent interacts with itself.
A more profound bidirectional interaction may be necessary for both the machines and humans to learn better. 
We envision a new paradigm of interactive machine learning, namely human-machine co-learning, where humans could learn from the model and the model could learn from humans as shown in Figure~\ref{fig:learningworkflows} (c). 
For example, in healthcare, clinicians may discover new knowledge or facts from the data through machine learning, and may help the machines to predict more accurately.  
Interactive learning could be applicable to many applications in the metaverse, in which humans interact with the model in both virtual and physical worlds, convey and discover new knowledge.
As a starting point, recent works \cite{wong2022assistq,lei2021assistsr} introduce AI assistants on smart glasses that can instruct novices in using a new device or learning a new skill. Yet, these models still do not form a bidirectional feedback loop between users and learning systems.
Such bidirectional interactive learning introduces new challenges to be tackled, namely novel interface design, model update, model interpretation and model intervention. 

\section{Conclusions}
\label{sec:conclusions}

The advancement in technologies has changed the way we live. In
the real world, we can participate in virtual games. In the world
of the virtual, we can shop, engage in strategic games that thrill
us and receive real-time information and acquire knowledge. The
merging of these two spaces will further enhance the user experience.
This paper has argued for the co-existence of the two spaces, not
as independent entities but as an integrated world where the two
spaces interact simultaneously, and users experience an
augmented world (either reality or virtuality) seamlessly. We have
presented several promising applications of the metaverse, and
discussed some research issues that the database community can
contribute.

In our discussion, we have focused primarily on the present; with
virtual space technology, time no longer ``bounds'' us - we can,
for example, be physically at a historical site experiencing
virtually an event that transpired in history on the exact spot
that we are standing; likewise, we can have a virtual futuristic
view of the current location.

As researchers, we look forward to the
exciting challenges in this field, and encourage members of our community
to join us. Perhaps, by 2030, we will experience the metaverse
as end-users and be brought ``back to the future''!


%
\bibliographystyle{abbrv}
\bibliography{sigproc}  
\balance
\end{document}